\documentclass[11pt]{article}
\usepackage{moriond}

\def\be{\begin{equation}}
\def\ee{\end{equation}}
\def\bea{\begin{eqnarray}}
\def\eea{\end{eqnarray}}

\newcommand{\Photo}{}

\begin{document}
\vspace*{1cm}
\title{ TOP-DOWN BEYOND THE STANDARD MODEL REVIEW\footnote{Talk presented
at Rencontres de Moriond EW2013,La Thuile, march 2-9, 2013.} }

\author{ Emilian DUDAS}

\address{Centre de Physique Th\'eorique, Ecole Polytechnique \\
91128 Palaiseau, France}

\maketitle\abstracts{
The recent discovery of the Standard Model boson (SMS) and direct searches place
new constraints and a new perspective on New Physics models. I mostly review supersymmetric model building, with special emphasizes on predictions of flavor 
models on superpartner spectra and inverted hierarchy models, mini-split models, very low-scale supersymmetry breaking predictions and some string theory inspired low-energy spectra.}

\section{Introduction}

Hierarchy problem guided ( or maybe misguided, depending on its own perspective), the physics beyond the Standard Model for the last thirty years. Traditional solutions fall into three categories: \\
- Low-energy supersymmetry  with superpartner masses in the TeV range $M_{\rm SUSY} \sim $ TeV. \\
- Strong dynamics like technicolor, Randall-Sundrum models, composite SMS models. \\
- Low-scale (TeV) strings /quantum gravity with or without supersymmetry
$M_{\rm SUSY} \sim M_{*} \sim $TeV. \\
Notice that in string theory the scale of supersymmetry breaking is not really
predicted to be in the TeV range and it could be much higher. Is is even possible that
$M_{\rm SUSY} \sim M_s \sim 10^{16}- 10^{17} \ $GeV, see the talk of A. Sagnotti \cite{augusto}. 

Starting in reverse order, flat extra dimensions provide spectacular low-energy physics: (sub)mm size gravitational (perpendicular) dimensions, TeV-size and possibly unification of gauge couplings from parallel dimensions, Kaluza-Klein dark matter,
etc. Current constraints from micro-gravity experiments set limits on perpendicular
dimensions $R_{\perp} < 0.02$ mm, whereas direct searches in colliders and 
indirect precision tests set the current limits on parallele dimensions $
R_{||}^{-1} > 1.5-2$ TeV \cite{takuda}. Due to lack of time, I will not discuss further flat extra dimensional models in this talk. 

The second solution to the hierarchy problem, strong dynamics, has its modern
incarnation in holographic models of the Randall-Sundrum type, with a Planck or ultraviolet (UV) brane and a TeV or infrared (IR) brane, with Standard Model states living in the bulk, but localized in the
fifth dimension by the various profiles of their wave functions. There is a conjectured holographic dictionary
\cite{tony} inspired by the AdS/CFT correspondence : \\
-5d states localized towards the TeV/IR brane are composite from a 4d viewpoint.
For example the 5d KK states are interpeted as resonances of a four-dimensional
strongly-coupled  theory. \\
- 5d states localized on the Planck/UV brane are interpreted as elementary 
states from a 4d perspective. \\
In such a framework, geometric localization leads to flavor structure. Current
limits from electroweak precision tests and flavor changing neutral current (FCNC)
effects put bounds on the IR scale of the order of $\Lambda_{IR} > 3 $ TeV.
A more severe bound $\Lambda_{IR} > 10 $ TeV arises if there is CP violation in the 
Yukawa sector. Most of the recent activity in this field was focused on composite
models for the scalar model boson \cite{azatov}, in which gauge symmetry is typically enhanced in the bulk to a higher one, the minimal example being based on the gauge group $SU(3) \times SO(5) \times
U(1)'$. In this example, the gauge symmetry is broken by boundary conditions to
the Standard Model on the Planck brane and to $SU(3) \times SO(4) \times
U(1)'$ on the TeV/IR brane. The $SO(4)$ factor on the IR brane contains the
custodial symmetry, which will survive as an approximate global symmetry. The SMS
is the fifth component of a gauge boson and is a pseudo-goldstone of the coset
$SO(5)/SO(4)$. Since it is localized on the IR brane, it behaves as a composite 
state. The lightest KK states in the model are colored fermions with electric charges
$-1/3,2/3$ and $5/3$, with masses between $0.5$ and $1.5$ TeV. The electrically charged state with charge $5/3$ decays mainly into $W^+ t \to W^+ W^+ b$, giving
a pair of same sign leptons in the final state.

\subsection{SUSY hints from LHC searches and BEH scalar mass }\label{subsec:prod}

LHC direct supersymmetry searches, the mass and the couplings of the recently discovered Standard Model scalar set new limits on superpartner masses for simple (simplified)
supersymmetric models \cite{marrouche}.  I think it is fair to say that popular 
models like minimal supergravity (mSUGRA), constrained Minimal Supersymmetric 
Standard Model (CMSSM) or minimal gauge mediation with TeV superpartner masses have some difficulties in accomodating the experimental data in a natural way \cite{marcela}. However, from a ultraviolet
(UV) point of view (supergravity,string theory) these models are rather unpopular,
i.e. they are difficult to obtain in specific string models with broken supersymmetry and which address flavor problems and moduli stabilization.
It is therefore important to theoretically propose and analyze and to experimentally search for non-minimal supersymmetric models. In what follows we display some non-minimal constructions; the first of them is motivated by flavor models for fermion mass hierarchies, the second by models of moduli stabilization, the third ones
by pushing to the extreme lowest values the scale of supersymmetry breaking and the last one is inspired by local models in recent F-theory constructions. 

\section{Inverted hierarchy / Natural SUSY models}

One old possibility \cite{ckn} which became popular recently because of LHC constraints on superpartner masses is
that of inverted hierarchy or, in its more extreme version, natural SUSY models. In such scenarios, the third generation squarks and gluinos have masses in the TeV range, in particular stops are light. On the other hand, the first two generation squarks are much heavier, typically $10-15$ TeV. They affect little however the
tuning of the electroweak scale, since their contribution to the electroweak v.e.v
is multiplied by their corresponding Yukawa couplings.  
Inverted hierarchy was invented in order to ease the FCNC and CP constraints
in supersymmetric models. Early ideas did invoke horizontal non-abelian symmetries for explaining fermion mass hierarchies  like $U(2)$  under which first two generations transform as a doublet, whereas the third generation is a singlet \cite{ckn}. Whereas $U(2)$  models do explain the difference and therefore can accomodate an hierarchy between the first two and the third generation of scalars, they do not actually predict it. To our knowledge, the first class of models 
in which the inverted hierarchy is really predicted are supersymmetric generalisations of abelian flavor models of the Froggatt-Nielsen
type \cite{fn}. These models contain an additional abelian gauge symmetry $U(1)_X$ under which the three fermion generations have different charges (therefore the name horizontal or flavor symmetry), spontaneously broken at a high energy scale by the
vev of (at least) one scalar field $\Phi$, such that $\epsilon = \langle \Phi \rangle / M << 1$ , where $M$ is the Planck scale or more
generically the scale where Yukawa couplings are generated.  
The order of magnitude of the quark Yukawa matrices in such models is given by
\be
h_{ij}^U \ \sim \ \epsilon^{q_i + u_j + h_u} \quad , \quad h_{ij}^D \ \sim \ \epsilon^{q_i + d_j + h_d} \ , \label{abelian1}
\ee
where $q_i$ ($u_i,d_i,h_u,h_d$) denote the $U(1)_X$ charges of the left-handed quarks (right-handed up-quarks, right-handed down-quarks,
$H_u$ and $H_d$, respectively). Quark masses and mixings in the simplest models are given as
\bea
&& \frac{m_u}{m_t} \sim \epsilon^{q_{13}+u_{13}} \quad , \quad \frac{m_c}{m_t} \sim \epsilon^{q_{23}+u_{23}} \quad , \quad \frac{m_d}{m_b} \sim \epsilon^{q_{13}+d_{13}} \quad , \quad \frac{m_s}{m_b} \sim \epsilon^{q_{23}+d_{23}} \ , \nonumber \\
&& \sin \theta_{12} \sim \epsilon^{q_{12}} \quad, \quad \sin \theta_{13} \sim \epsilon^{q_{13}} \quad , \quad 
\sin \theta_{23} \sim \epsilon^{q_{23}} \ . \label{abelian2}
\eea  
A successful fit of the experimental data requires larger charges for the lighter generations
\be
q_1 \ > q_2 \ > q_3 \quad , \quad u_1 \ > u_2 \ > u_3 \quad , \quad d_1 \ > d_2 \ > d_3 \ ,  \label{abelian3}
\ee
one simple example, using as small parameter the Cabibbo angle 
$\epsilon = \sin \theta_c$, being defined by the charges 
\be
q_1 = 3 \ , \ q_2 = 2 \ , \ q_3 = 0 \ , u_1 = 5 \ , \ u_2 = 2 \ , \ u_3 = 0 \ , d_1 = 1 \ , \ d_2 = 0 \ , \ d_3 = 0 \ .  \label{abelian4} 
\ee
Scalar soft masses in abelian flavor models are typically of the form
\be
m_{ij}^2 \ = \ X_i \langle D \rangle \ + \ c_{ij} \ \epsilon^{|q_i-q_j|} 
\ (\frac{F}{M})^2 \ , \label{abelian5} 
\ee 
where $X_i \langle D \rangle$ are D-term contributions for the scalar of charge $X_i$, whereas  the last term denote the F-term contributions, also constrained by the
abelian symmetry.  
The D-term contributions were argued to be naturally generated  in effective string models, to be positive and, in certain circumstances, to be dominant over the F-term contributions \cite{bddp}. It is then clear from (\ref{abelian3}),(\ref{abelian4}) that precisely because the first generations of fermions are lighter than the third one, the corresponding scalars are {\it predicted to be heavier} \cite{carlos}. 

Abelian and non-abelian flavor models are complementary in one respect: whereas abelian models naturally predict the inverted hierarchy, which is just an option in the non-abelian case,  
they do not generically predict approximate degeneracy among the first two generations, unlike their non-abelian cousins.
This leads to possible tension with FCNC constraints, which were analyzed in some details in the literature. This means that inverted hierarchy models do generically predict  $m_{Q_i} \not= m_{U_i} \not= m_{D_i} $. 
Since the first two generations are very heavy, we could expect much larger RGE effects than in the universal case $m_{Q_i}= m_{U_i}= m_{D_i}$. Indeed, the RGE's
of all scalar soft masses and in particular of the third generation of squarks and of the Higgs scalars depend to some extend of
the combination 
\be
S \ = \ Tr (Y m^2) \ = \ m_{H_u}^2-m_{H_d}^2+\sum_{i=1}^3[m_{Q_i}^2-2m_{U_i}^2+m_{D_i}^2-m_{L_i}^2+m_{E_i}^2]  \ ,  \label{abelian7} 
\ee 
which is zero at high-energy in the universal case, where the trace is over the whole spectrum of MSSM states. Interestingly enough, in abelian flavor models with D-term dominance of the type discussed here, the quantity S is approximately equal to
\be
S \ = \ Tr (Y X) \  \langle D \rangle \ .  \label{abelian8} 
\ee
However, $Tr (Y X)$ has to vanish (or to be very small) for phenomenological reasons, as argued in various papers. The running of soft terms and the fine-tuning of the
electroweak scale was discussed in \cite{marcin}. It was noticed there that
there is a region in parameter space where the stop becomes light and the stop mixing
becomes large due to the RG effect coming from the first two generation squarks. Indeed, due to their heavy mass, at two-loops they affect significantly the stop running and have  the tendency to render the stop light and even tachyonic.
  
\section{Mini-split SUSY models}

Mini-split models \cite{minisplit} are version of split supersymmetry \cite{split}, with scalar and higgsino masses in the mass range $30-500$ TeV and gaugino masses in the TeV range, due to a loop suppression. 
Natural realizations of mini-split scenario arise in "pure gravity mediation" \cite{yanagida} or
"strong moduli stabilization" models \cite{dlmmo}, in which scalar masses are fixed by the gravitino mass $m_0 \sim m_{3/2}$, whereas gaugino masses and A-terms are fixed
by anomaly mediation
\be
M_{1/2}^a \ = \ \frac{b_a g_a^2}{16 \pi^2} m_{3/2} \ . 
\ee
Models with strong moduli stabilization were initially proposed in order to
solve cosmological problems like vacuum destabilization during inflation
and moduli problem in models of moduli stabilization in string theory \cite{kl}. In such models, moduli masses $T$ and the mass of the field breaking supersymmetry $S$
are much higher than the gravitino mass $m_{3/2}$, which is in the range $30-500$ 
TeV. Moduli $T$ have a very small contribution to supersymmetry breaking, which is almost entirely provided by $S$. If 
$S$ has a small coupling to MSSM fields in the Kahler potential and superpotential, then the mini-split spectrum with scalar masses close to the gravitino mass and gaugino masses given by anomaly mediation arises naturally. The LSP in this case is
the wino, as in anomaly mediation. 

In such models, there are strong correlations between: \\
- the SMS mass and the gravitino mass. For a fixed Standard Model mass, there is an upper limit on the scalar superpartner masses and therefore on the gravitino mass. For example, an SMS mass of $125$ GeV implies a limit of $50-100$
TeV on scalar masses. 
 \\
- the relic density of wino LSP and the gravitino mass, which determine an upper
bound $m_{3/2} < 650 $ TeV. In this upper limit case, the Higgs mass is on the heavy
side, around $128.5$ GeV, which is by now far from the central SMS
mass value. For lower scalar
masses, compatible with the central value for the SMS mass, the relic density of the 
LSP winos is too small compared to the needed value 
$\Omega_{\chi} h^2 \simeq 0.11$. In this case, one needs other options to increase
relic density. One option is simply another dark matter component, for example
axions. The second logical option is a non-thermal production of LSP through decays
of moduli fields or gravitinos.  

\section{Low-scale SUSY breaking dynamics}\label{subsec:fig}

Spontaneous breaking of global supersymmetry leads, through the
Goldstone theorem, to the existence of a massles fermion, the goldstino. In its gauged version (supergravity), analogously to the Higgs mechanism, the goldstino provides the longitudinal components and is absorbed by the gravitino, which therefore becomes 
massive. The goldstino is part of a supersymmetric multiplet, which can be chiral or vector. In what follows we consider goldstino to be part of a chiral multiplet $X = (x, G, F_X) $, where its scalar superpartner is called sgoldstino in what follows. The sgoldstino mass $m_x$ depends on the microscopic theory of supersymmetry breaking.
In a SUSY theory well below the scale of SUSY breaking $E << \sqrt{f}$, SUSY is  non-linearly realized. 
For low scale of supersymmetry breaking $\langle |F_X| \rangle \simeq f << m_{sparticles}^2$, where $m_{sparticles}$ is the typical mass scale of superpartner masses, there is always one light fermion in the effective theory, the goldstino $G$ or more precisely the gravitino which couples to matter through its helicity $1/2$ components, of mass
\be
m_G \sim \frac{f}{M_P} \ . \label{ls1}
\ee
In the decoupling limit $M_P , m_x \rightarrow \infty$, with fixed scale of supersymmetry breaking $f$, the transverse polarizations of the  gravitino decouple, whereas its longitudinal component (goldstino)  couplings scale as $1/f$.

There are three qualitatively different cases of goldstino couplings to matter, depending on the masses of superpartners and sgoldstino versus the energy of the process : \\
i) Non-SUSY matter spectrum, for example the Standard Model coupled to the goldstino,
if
\be
E \ << \ m_{sparticles} \ , \ m_x \ , \ \sqrt{f} \ . \label{ls2}
\ee
In this case, there is a non-linear realization of supersymmetry in the matter sector. This is the straightforward generalization of the original Volkov-Akulov lagrangian. All models of supersymmetry breaking at energies below the scale of supersymmetry breaking enter into this category. If one wants a low-scale of supersymmetry breaking $\sqrt{f} \sim $ 5-10 
TeV, one expects the underlying microscopic degres of freedom, superpartners for
field theory models or string states for string theory, to have similar masses. Explicit realizations of models in this class include string models with non-linear supersymmetry with low string scale $M_s \sim $ TeV \cite{bsb}. This is not the regime
that will be discussed in what follows. \\
ii) SUSY matter multiplets like in MSSM: quarks-squarks, gauge fields-gauginos, etc,
but with non-linear supersymmetry in the goldstino multiplet sector, i.e. heavy sgoldstino 
\be
m_{sparticles} \sim E << \sqrt{f} \ , m_x \ . \label{ls3}
\ee
In this case, the matter sector has a linearly relized supersymmetry,  coupled to the goldstino. This is one  energy regime we will consider in what follows, dubbed
non-linear MSSM \cite{adgt}. This framework 
leads, in addition to the standard MSSM soft terms and known goldstino couplings, to new MSSM couplings, and in particular to correction to the SMS potential.

iii) Supersymmetric multiplets with linearly realized supersymmetry, for energies
 such that all superpartners and the goldstino are accessible 
\be
E \ \sim \ m_{sparticles} \ , \ m_x \  < \ \sqrt{f} \ . \label{ls2}
\ee
This regime corresponds to standard linear realization of supersymmetry in all sectors, with non-renormalizable couplings of the supersymmetry ($X$) breaking sector to the MSSM sector. The origin of
these couplings should be related to strong dynamics at low-energy, coupling the
supersymmetric breaking sector to the observable one. 
  
In both cases ii) and iii) above, not much is known about the explicit construction
of models with low fundamental scale. The minimal ingredients for explicit construction of such models should include a supersymmetry breaking sector at TeV low-energy and a mediation of supersymmetry breaking via strongly-coupled messengers. This is needed in order to overcome
the usual lower-bound on supersymmetry breaking scale $\sqrt{f} > 50-100$ TeV in
gauge mediation models, based on perturbative loop-induced soft terms.

Let us start with the case ii) above, in which supersymmetry is non-linearly realized only in the goldstino sector. In what follows  we are using the superfield approach of Rocek \cite{rocek}, in which
the Goldstino $G$ can be described by a  chiral superfield $X$, subject to the
superfield constraint
\be
X^2 \ = \ 0 \ . \label{ls3}
\ee
The constraint is solved by
\be
X \ = \ \frac{G G}{2\, F_X} +\sqrt 2\,\, \theta G
+\theta\theta\,\,F_X \ , \label{ls4}
\ee
where the auxiliary field $F_X$ is  to be eliminated via its field equations.

Usually we parameterize SUSY breaking in supersymmetric extensions of the Standard Model by coupling matter fields to a spurion with
no dynamics $S \ = \ \theta^2 m_{soft} $. 
The main difference in the context of the non-linear MSSM is the replacement of the spurion with a dynamical constrained superfield $S \rightarrow \frac{m_{soft}}{f} \ X$. This reproduces the MSSM soft terms, but it contains simultaneously the goldstino
couplings to matter. Moreover, it adds new dynamics. The fact that $F_X$ is a dynamical auxiliary field, determined as usual through its algebraic field equations, generates new couplings : 
\be
- {\bar F}_X = f + \frac{B}{f} h_1 h_2 + \frac{A_u}{f} {\tilde q} {\tilde u} h_2 + \cdots \label{ls6}
\ee
The formalism  contains in a very compact, superfield form, the goldstino couplings to matter.
The one-goldstino couplings are on-shell equivalent to the ones based on the standard supercurrent coupling of the goldstino  $\frac{1}{f} \partial_{\mu} G \, J^{\mu}$. For processes in which some particles are off-shell, comparison with standard approach was checked in some instances but,
to my knowledge, not completely.  

In this formalism, all couplings to the Goldstino are proportional to soft-terms.
The lagrangian is schematically
\be
{\cal L} \ = \ {\cal L}_{MSSM} + {\cal L}_X + {\cal L}_m + {\cal L}_{AB} + {\cal L}_g \quad , \quad {\rm where} \label{ls7}
\ee
\bea
&&{\cal L}_H = \sum_{i=1,2} \frac{m_i^2}{f^2} \int d^4 \theta \ X^{\dagger} X \ H_i^{\dagger} e^{V_i} H_i \ , \nonumber \\
&&{\cal L}_m = \sum_{\Phi} \frac{m_{\Phi}^2}{f^2} \int d^4 \theta \ X^{\dagger} X \Phi^{\dagger} e^{V} \Phi \ , \ \Phi = Q,U_c,D_c,L,E_c \ , \nonumber \\
&&{\cal L}_{AB} =  \frac{B}{f} \int d^2 \theta \ X H_1 H_2 + ( \frac{A_u}{f} \int d^2 \theta \ X Q U_c H_2 + \cdots)  \ , \nonumber \\
 && {\cal L}_{g} = \sum_{i=1}^3 \frac{1}{16\, g^2_i\,\kappa}
\frac{2\,m_{\lambda_i}}{f} \int d^2\theta
\,X \,\mbox{Tr}\,[\,W^\alpha\,W_\alpha]_i +h.c. \quad . \label{ls8}
\eea
This lagrangian is still a parametrization and not an explicit model of supersymmetry breaking. The origin
of soft terms is not specified and their values are just parametrized, like in MSSM
with a spurion. But such lagrangian contains more than the MSSM lagrangian with soft
terms. In addition to goldstino couplings,  
matter terms coming from solving for $F_X$ are new; they do not come from a standard Volkov-Akulov non-linear supersymmetry realization prescription. The most interesting
example of a new coupling is the scalar potential, which is modified compared to MSSM and is given by:
\bea
&& V=
\big(\vert \mu\vert^2+m_1^2\big)\,\,
\vert h_1\vert^2+
\big(\vert \mu\vert^2 +m_2^2\big)
\vert h_2\vert^2
+ ( B\,h_1.h_2 + {\rm h.c.}) \nonumber \\[2pt]
&& + \frac{g_1^2+g_2^2}{8}\,\Big[\vert h_1\vert^2-\vert h_2\vert^2\Big]^2
+\frac{g_2^2}{2}\,\vert h_1^\dagger\,h_2\vert^2 
 +   \frac{1}{f^2}\,\Big\vert m_1^2\,\vert h_1\vert^2+m_2^2\,\vert h_2\vert^2+
B\,h_1.h_2\Big\vert^2  \ .  \label{ls9}
\eea
The last term  in (\ref{ls9}) is new compared to MSSM. It contains new quartic couplings not related to gauge couplings like in the usual MSSM potential,
but rather related to the soft terms and the scale of supersymmetry breaking. It is generated by integrating out the sgoldstino multiplet and its
physical interpretation should be related to new couplings of the Higgs multiplet to the (low-scale) supersymmetry  breaking sector.

It was shown \cite{adgt} that this frawmework can raise the Standard Model boson
mass up to the ATLAS and CMS values by the tree-level contributions of the goldstino
auxiliary field displayed in (\ref{ls9}). 
On the other hand, the one-goldstino couplings to the MSSM fields that one finds contain the usual supercurrent couplings. This is obtained in the  setup
(\ref{ls8}) containing MSSM plus the minimal set of operators needed to parameterize the soft-breaking terms \cite{adgt}. One can show that the effect of additional higher-dimensional/derivative operators is to correct existing MSSM  couplings $\lambda$ in the following generic way \cite{dpt},
\be
\lambda \ = \ \lambda_{MSSM} \left( 1 + \sum_n c_n \left(\frac{M_{SUSY}}{\sqrt{f}}\right)^n \right) \ , \label{intro1}
\ee
where $M_{SUSY} \sim M_{sparticles}$ is the scale of supersymmetry breaking in the observable sector, generating sparticle masses. Since by consistency $M_{sparticles} < \sqrt{f}$, the corrections (\ref{intro1}) to an existent tree-level MSSM coupling are small. Some couplings however, which are loop-generated or
small at tree-level can receive important corrections, such as the SMS self-coupling
or the SMS decay into two photons $h\rightarrow \gamma\gamma$.
The renormalizable tree level SMS couplings can be parametrized as
\be
\label{Ltree}
{\cal L}_{ren} =  -c_t \frac{m_t}{v}h \,t \,\bar{t}-c_c \frac{m_c}{v}h \,c \,\bar{c}-c_b \frac{m_b}{v}h \,b \,\bar{b}-c_\tau \frac{m_\tau}{v}h \,\tau \,\bar{\tau} 
+  c_Z \frac{m_Z^2}{v}h \,Z^{\mu} \,Z_\mu +  c_W \frac{2m_W^2}{v}h \,W^{+\mu} \,W^{-}_\mu \ . 
\ee
In the MSSM decoupling limit: $c=1$ and the loop contributions $c^{\mathrm{loop}}$ equal the SM ones. In case iii) above with light sgoldstino scalar, there is an interesting phenomenon,  a sgoldstino-higgs mass mixing, which leads to possible enhancement in $h\rightarrow \gamma\gamma$ \cite{christoffer,dpt}.
It comes from 
\be
{\cal L}\supset x\left(-{m_i^2\over f^2}F_X^\dag \, h_i^\dag F_i+{B\over f} (F_1h_2+h_1F_2)-{M_a\over 4f}(F^{k\, \mu\nu}F^k_{\mu\nu})_a\right) + h.c.
-|x|^2 \left({m_i^2\over f^2}|F_i|^2+m_X^2\right) \ .  \label{ls10}
\ee
If the sgoldstino $x$ is heavy we can use its equations of motion at zero-momentum  
to integrate it out. We obtain
\be
-{M_a\over 4m_X^2f^2}(F^{k\, \mu\nu}F^k_{\mu\nu})_a\left( m_i^2 h_i^\dag \,F_i+B(F_1h_2+h_1F_2)\right)+h.c.~.
\ee
This generates an  effective interactions between the SMS $h$ and the gauge  field strengths. Then
 \begin{equation}
\label{coeff}
c_\gamma = c_\gamma^{\mathrm{loop}} +c_\gamma^{\mathrm{sgold}} \ , \ c_g = c_g^{\mathrm{loop}}+c_g^{\mathrm{sgold}} \ , \
c_{Z\gamma} = c_{Z\gamma}^{\mathrm{loop}}+c_{Z\gamma}^{\mathrm{sgold}} \ ,
 \end{equation}
where,
 \begin{eqnarray}
 \label{coeffB}
c_\gamma^{\mathrm{sgold}} &=& -\frac{4\pi\,v^2\mu}{f^2m_X^2\alpha_{\mathrm{EM}}} (M_{1}\cos^2\theta_w+M_{2} \sin^2\theta_w) \,\Delta \ , \nonumber \\
c_{Z\gamma}^{\mathrm{sgold}} &=& -\frac{4\pi\,v^2\mu\cos\theta_w\sin^2\theta_w}{f^2m_X^2\alpha_{\mathrm{EM}}}(M_{1}-M_{2})\,\Delta \ 
\ , \ 
c_{g}^{\mathrm{sgold}} =-\frac{6\pi\,v^2\mu}{f^2m_X^2\alpha_{\mathrm{S}}}\, M_{3} \,\Delta  \ .
 \end{eqnarray}
 The factor $\Delta$ is written explicitly in \cite{dpt} and equals 
 $\Delta \to \mu^2 \sin 2 \beta$ in the MSSM decoupling limit. 
We can then use the experimental bound on the gluino mass, which enters the $c_{g}^{sgold}$ to estimate how much the Higgs couplings to $\gamma\gamma$ and $ Z\gamma$ can be enhanced. 
If we do not want gluon fusion to deviate from SM value by more than around 30\%, i.e. \mbox{$|c_{g}^{sgold}| \leq 0.14\cdot |c_g^{SM}|$}, then
there is a lower limit on the supersymmetry breaking scale. By combining this with the expression for $c_{\gamma}^{sgold} $ gives the bound
$ \left| c_\gamma^{sgold} \right|
\leq 1.37 \, \left| \frac{M_{12}}{M_{3}} \right|$, 
where $M_{12}=M_{1}\cos^2\theta_w+M_{2} \sin^2\theta_w$. Assuming the signs of $\mu$ and $M_{12}$ are such that the sgoldstino mixing contribution is constructive, this implies
\begin{equation}
\label{gammagamma}
\frac{\Gamma_{h\gamma\gamma}}{\Gamma_{h\gamma\gamma}^{\mathrm{SM}}}=\left| \frac{c_\gamma}{c_\gamma^{\mathrm{SM}}} \right|^2 \ \leq \ \left| 1 + 0.21 \frac{M_{12}}{M_{3}} \right|^2\,.
\end{equation}

The result (\ref{gammagamma}) suggests reasonable enhancement or suppression of the
SMS branching ratio $h \to \gamma \gamma$ of the order $10-20$ $\%$ , although smaller
deviations are expected for the (rather standard) case of gluinos heavier than the binos and the winos. 
 
\section{String and F-theory inspired SUSY spectra}\label{sec:plac}

Recently there was an intense activity in constructing F-theory models of particle
physics, especially in building $SU(5)$ GUT models with additionial $U(1)$ gauge
symmetries \cite{eran}. In such models, the GUT gauge group is localized on a D7 brane
wrapping a four space called the GUT divisor. Typically there are magnetic type fluxes in the internal space along the hypercharge generator and in the additional $U(1)$ gauge factors. Hypercharge flux is needed to break $SU(5)$ down to the Standard Model
gauge group, whereas $U(1)$ fluxes generate the chirality necessary in order
to reproduce the MSSM spectrum. The internal volume of the GUT brane is described by
a modulus field, called GUT modulus in what follows. The hypothesis made by the
recent papers  \cite{madrid} is that this modulus is responsible for breaking supersymmetry. In this case, scanning over one parameter flux, they found that
soft terms generated at the string scale satisfy naturally the relations:
\be
M_{1/2} \ = \ \sqrt{2} \ m_0 \ = \ - \frac{2}{3} A \ = \  - B \ . 
\ee
In particular, $A \simeq - 2 m_0$  and, after running from the fundamental string scale down to the TeV scale, this pattern of soft masses generate a nearly maximal stop mixing needed in order to increase the Higgs mass to $125$ GeV with relatively light stop masses. This example shows that it is possible to get a nearly maximal stop mixing naturally from a microscopic theory like string theory.

\section*{Acknowledgments}

 I would like to thank the organizers for their kind invitation to give a talk
in a nice and pleasant environment. Many thanks to  Marcin Badziak, Andrei Linde, Yann Mambrini, Azar Mustafayev, Marek Olechowski,  Keith Olive, Christoffer Petersson, Stefan Pokorski and Pantelis Tziveloglou for collaboration on the different works on which this
talk is partially based. I would like to also thank the Galileo Galilei Institute for Theoretical Physics for the hospitality and the INFN for partial support. This  work was supported in part by the European ERC Advanced Grant 226371 MassTeV, the French TAPDMS ANR-09-JCJC-0146 and the contract PITN-GA-2009-237920 UNILHC.


\end{document}